\documentstyle[epsfig, twocolumn]{venice97}

  \newcommand{\teff}{\mbox{\,\em T$_{\rm eff}$}}         
  \newcommand{\logg}{\mbox{\,log $g$}}                   
  \newcommand{\Msolar}{\mbox{\,$\rm M_{\odot}$}}        

\begin{document}

\setlength{\parindent}{0pt}
\setlength{\parskip}{ 10pt plus 1pt minus 1pt}
\setlength{\hoffset}{-1.5truecm}
\setlength{\textwidth}{ 17.1truecm }
\setlength{\columnsep}{1truecm }
\setlength{\columnseprule}{0pt}
\setlength{\headheight}{12pt}
\setlength{\headsep}{20pt}
\pagestyle{veniceheadings}

\vspace*{0.5cm}
\title{\bf MASSES AND GRAVITIES OF BLUE HORIZONTAL BRANCH 
(BHB) STARS REVISITED}

\author{{\bf U.~Heber$^1$, S.~Moehler$^1$, I.N.~Reid$^2$} \vspace{2mm} \\
$^1$Astronomisches Institut, Universit\"at Erlangen-N\"urnberg, 
Dr. Remeis-Sternwarte, Bamberg, Germany\\
$^2$Palomar Observatory, Pasadena, U.S.A.}

\maketitle

\begin{abstract}

Previous spectroscopic analyses of Blue Horizontal Branch (BHB) stars in 
six globular clusters revealed too low masses in four clusters when 
compared to canonical evolutionary theory, while the masses of the BHB stars 
in NGC\,6752 and M\,5 are found to be consistent with theory.  
We recalculated BHB star masses using new cluster 
distances derived by Reid (1997a,b) from HIPPARCOS parallaxes of local 
subdwarfs by main sequence fitting. The new distances are larger than 
previous estimates resulting in larger masses for the BHB stars. Since the 
increase in distance is small for NGC\,6752 and M\,5, the agreement
with predicted masses persists. For M\,15 and M\,92 the masses now come into 
good agreement with theoretical predictions, while for NGC\,288 and 
NGC\,6397 the mass deficit is reduced but the BHB star masses remain 
slightly too low. Previous spectroscopic analyses also highlighted the 
problem of too low gravities for some BHB stars. The gravities and absolute 
magnitudes of BHB stars are revisited
in the light of new evolutionary Horizontal Branch models.
  \vspace {5pt} \\


  Key~words: Globular clusters; Blue Horizontal Branch stars; masses; 
gravities.

\end{abstract}

\section{INTRODUCTION}

%

Recent spectroscopic investigations of blue horizontal branch (BHB) stars in 
the globular clusters NGC\,288, M\,5, M\,92, NGC\,6397, NGC\,6752 and M\,15
(\cite{cro88}; 
\cite{moe95}, \cite{moe97}, and
\cite{deb95}, hereafter CRO88 and papers I,III,II) 
showed that only in M\,5 are the BHB stars an almost perfectly match the ZAHB 
prescription, 
both in gravity and in mass. The hottest HB stars (\teff\ $>$ 20000~K 
and \logg\ $>$ 5.0) are termed extreme 
HB stars or sdB stars since they spectroscopically resemble the sdB 
stars in the field very well (see \cite{heb92}).
Besides M\,5, NGC\,6752 comes closest to the theoretical 
prediction at least for the hot end of the HB (sdB stars).
Significant discrepancies between observational 
results and predictions by classical theory resulted for the other clusters
and two main problems emerged:
\begin{enumerate}
\item The BHB stars (i.e. stars with 10,000~K $<$ 
\teff\ $<$ 20000~K and \logg\ $<$ 5) showed {\bf too low gravities} 
compared to standard evolutionary theory (CRO88, I). 
On the other hand cooler BHB stars as well as sdB stars 
were in good agreement with 
canonical tracks (II, III) in the (\teff,\logg) diagram.
This is illustrated in 
Figure 4.

\item All BHB stars (except for M\,5) had masses (on average) 
significantly below the canonical values of 0.5 -- 0.6~\Msolar 
({\bf too low masses}, I,II). 
The masses of the hottest HB stars (sdB stars) in NGC~6752, however, 
scatter around a mean value of 0.49~\Msolar,
very close to what is expected from theory (0.5~\Msolar).
(III). 
\end{enumerate}

The above mentioned papers also discussed a variety of possible 
explanations. The main problem turned out to be that it was not possible to 
find one solution that explained the too low gravities and masses at the 
same time, while keeping the ``correct'' masses and gravities
of the sdB stars and the gravities of the cooler BHB stars.


Recently published HIPPARCOS results (Reid 1997a,b, 
\cite{gra97}) as well 
as new evolutionary tracks (\cite{swe97}) 
shed new light on these problems 
and offer solutions to the observed discrepancies.  

\section{HIPPARCOS PARALLAXES OF LOCAL SUBDWARFS AND 
THE DISTANCES TO GLOBULAR CLUSTERS}

Reid (1997a,b) and Gratton et al. (1997) used HIPPARCOS 
parallaxes of nearby subdwarfs to redetermine the distances to a number of 
globular clusters, primarily 
using the abundance scales given by high resolution spectroscopy of 
cluster giants and field subdwarfs (\cite{cg97}, \cite{gra97}, \cite{axe94}).
Details of the fitting procedures can be found in the quoted papers. 
In Table 1 we give 
the new distances (together with the adopted reddenings and metallicities) 
for the globular clusters in question.
\begin{table*}
\begin{tabular}{|l|lll|lll|lll|}
\hline
Cluster & \multicolumn{3}{c|}{\cite{djo93}} &
\multicolumn{3}{c|}{Reid (1997a,b)} &
\multicolumn{3}{c|}{Gratton et al. (1997)}\\
 & [Fe/H] & (m-M)$_{\rm V,0}$ & E$_{\rm B-V}$ &
 [Fe/H] & (m-M)$_{\rm V,0}$ & E$_{\rm B-V}$ &
 [Fe/H] & (m-M)$_{\rm V,0}$ & E$_{\rm B-V}$ \\
\hline
NGC 288         & $-$1.40 & 14.62 & 0.03 & $-$1.07 & 15.00 & 0.01 & 
 $-1.11$ & 14.76 & 0.001 \\
NGC 5904 (M 5)  & $-$1.40 & 14.40 & 0.03 & $-$1.10 & 14.53 & 0.02 & $-$1.11 
 & 14.58 & 0.029 \\
NGC 6341 (M 92) & $-$2.24 & 14.38 & 0.02 & $-$2.24 & 14.93 & 0.02 & $-$2.17 
 & 14.83 & 0.013 \\
NGC 6397        & $-$1.91 & 11.71$^2$ & 0.18 & $-$1.82 & 12.25 & 
0.19 & & & \\
NGC 6752        & $-$1.54 & 13.12 & 0.04 & $-$1.42 & 13.17 & 0.02 & $-$1.40 
 & 13.20 & 0.028 \\
NGC 7078 (M 15) & $-$2.17 & 15.11 & 0.05 & $-$2.15 & 15.45 & 0.11 & & & \\
\hline
\end{tabular}
\begin{tabular}{l}
$^2$ De Boer et al. (1995) used a distance modulus of 12.0 for 
their mass determinations\\
\end{tabular}
\caption[]{The distances and reddenings derived from the new HIPPARCOS 
parallaxes compared to the old values.}
\end{table*}

As can be seen from Table~1 the new distance moduli are always larger than the 
old ones, in some cases by up to 0.55~mag (M\,92). 
The effects are most pronounced for the metal poor
 clusters M~15, M~92 and NGC~6397, 
whereas the intermediate metallicity clusters NGC~6752 and M~5 are almost 
unchanged. The new distance for NGC~288 derived by Reid (1997b) is 
somewhat larger than that derived by  Gratton et al. (1997).

\section{MASSES OF BHB STARS REVISITED}

Because the masses are derived from apparent magnitude, gravity and 
cluster distance, larger distances will result in 
larger masses for the stars observed in those clusters. In Figures 1--3 we plot 
the masses obtained with distance moduli listed by Djorgovski (1993) in 
comparison with the masses obtained from Reid's new distance moduli. 
Since the mass determination for an individual star has an uncertainty of at 
least $\pm$0.2\,dex, only the mass distributions are significant 
for such an comparison.   
 
Figure 1 compares the masses of the BHB stars calculated from the new 
distance scale with those obtained from the 
old one for 
the clusters M~15, M~92 and NGC~6397, for which the new distances are 
substantially larger. As can be seen the new masses for BHB stars in M\,92 
and M\,15 perfectly match the theoretical prescription. The scatter of the 
M\,15 masses is considerably larger than for M\,92 due to the lower 
accuracy of the gravities for the former cluster. 
Also 
the masses of the BHB stars in NGC~6397 are now closer to canonical 
predictions although there still remains a deficit. 

Figure 2 plots the BHB 
and sdB star masses for NGC\,6752 and M\,5, the clusters with the smallest 
changes in distance moduli. Since the distance changes are small, the 
corresponding changes in masses are also small and the overall agreement 
of spectroscopically determined masses with the theoretical predictions 
remains very good.  

Figure 3 plots the results for NGC~288. Since the new distance derived by  
Reid (1997b) is somewhat larger than the one derived by Gratton et al. (1997), 
we 
plotted masses calculated from both distance estimates in the bottom panel 
of Figure 3. 
As in the case of NGC\,6397 the resulting masses are still too low when 
compared to the canonical values, even for the largest distance modulus of 
Reid (1997b), but the mass deficit is greatly reduced.  
Note in passing that the metallicity of NGC~288 rests on only two red giant 
stars, and therefore the determination of its distance 
might be affected via the selection of calibrating subdwarfs. VandenBerg et 
al. (1990)
argue that the colour-magnitude diagram, and therefore the abundance is 
very similar to that of M\,5, which matches the Caretta \& Gratton 
measurements.

{\small
\begin{figure}[!ht]
   \begin{center}
    \leavevmode
\centerline{\epsfig{file=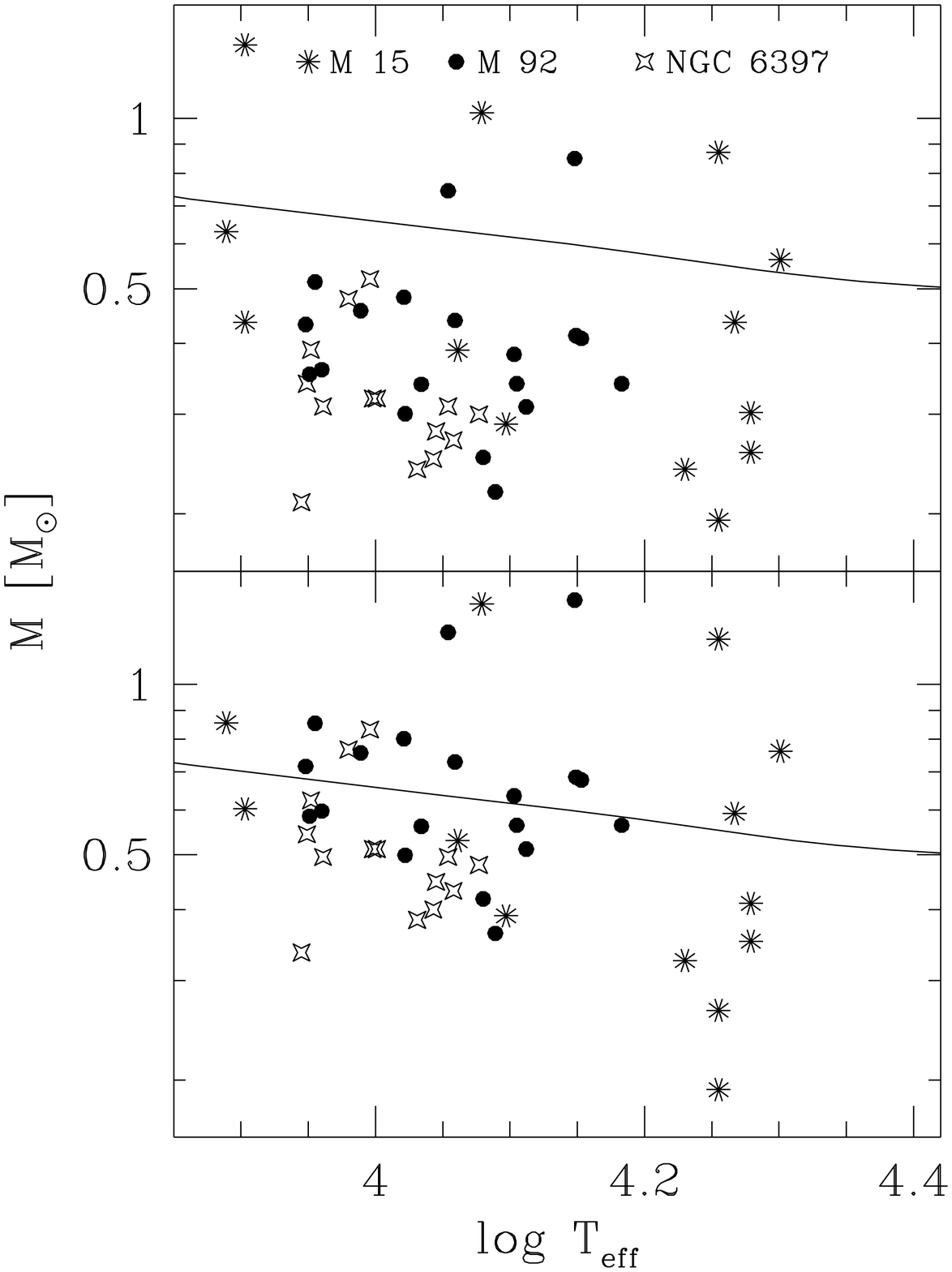,height=8.3cm}}
   \end{center}
\vspace*{1.1cm}
\caption[]{The masses of BHB and sdB stars in the metal poor globular clusters
M15, NGC\,6397 and M\,92. The full drawn lines give the theoretical 
prediction from the models of Dorman et al. (1993, [Fe/H=-2.26]) {\bf Top:}
The masses obtained with distance moduli as given by Djorgovski (1993).
{\bf Bottom:} masses obtained with the new distance moduli of Reid (1997a,b).}
\end{figure}
}

{\small
\begin{figure}[!ht]
   \begin{center}
    \leavevmode
\centerline{\epsfig{file=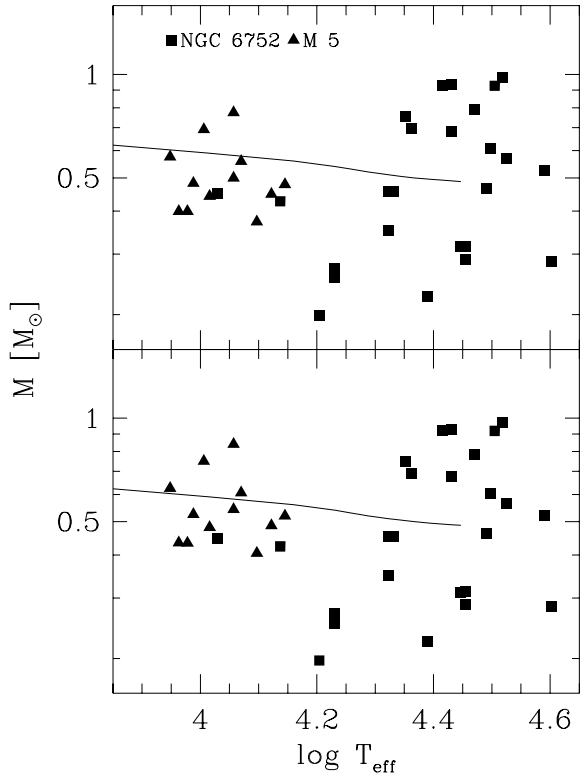,height=8.3cm}}
   \end{center}
\vspace*{1.1cm}
\caption[]{The masses of BHB stars in the intermediate metallicity 
globular clusters
M\,5 and NGC\,6752. The full drawn line gives the theoretical 
prediction from the models of Dorman et al. (1993, [Fe/H=-1.48]). {\bf Top:}
The masses obtained with distance moduli as given by Djorgovski (1993).
{\bf Bottom:} masses obtained with the new distance moduli of Reid (1997b).}
\end{figure}
}

{\small
\begin{figure}[!ht]
   \begin{center}
    \leavevmode
\centerline{\epsfig{file=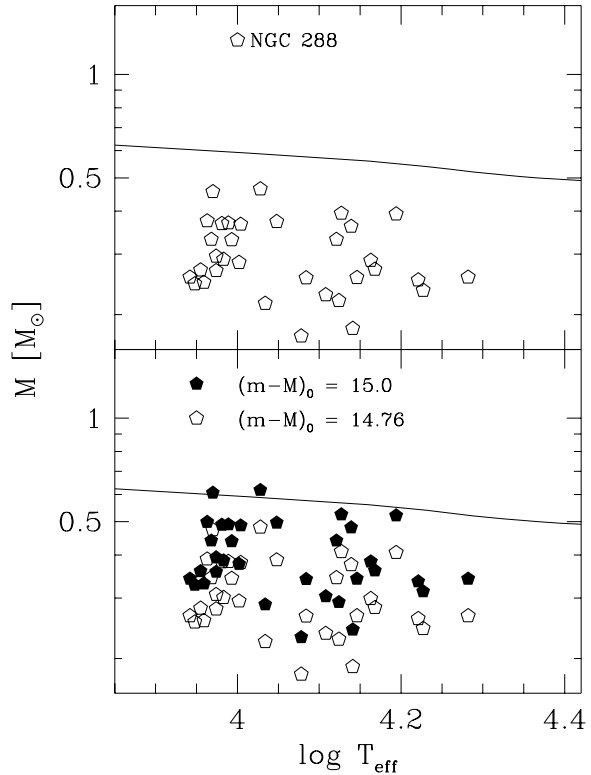,height=8.3cm}}
   \end{center}
\vspace*{1.1cm}
\caption[]{The masses of BHB stars in the metal poor globular clusters
NGC\,288. The full drawn lines give the theoretical 
prediction from the models of Dorman et al. (1993, [Fe/H=-1.48]). {\bf Top:}
The masses obtained with distance moduli as given by Djorgovski (1993)
{\bf Bottom:} masses obtained with the new distance moduli of Reid (1997b, 
full drawn) and Gratton et al. (1997, open symbols).}
\end{figure}
}


\section{GRAVITIES OF BHB STARS REVISITED}

As mentioned in the introduction not only the masses of the BHB stars
posed a problem, but also their gravities. On average the gravities for 
stars with \teff\ between 10000~K and 20000~K are too low, even when taking 
evolutionary effects into account. This problem is not affected at all by 
the distances of the objects so we have to look for another solution. 
Recently non-canonical  
models were calculated by Sweigart (1997) which take deep helium mixing
 into account. In Figure 4 we compare the physical 
parameters of the cluster stars to these new tracks.

It is obvious from Figure 4 that deep helium mixing could explain the low 
gravities found for effective temperatures between 10000~K and 20000~K,
while leaving the tracks for the sdB region virtually unchanged. However,
there is no indication for helium mixing in the cooler BHB stars and it is 
not readily clear, why mixing should be important only at sufficiently high 
HB temperatures.
On the other hand, stars with deep helium mixing tend to achieve bluer 
positions along the BHB, so there might be some correlation between the 
position on the BHB and the amount of helium mixing.

{\small
\begin{figure}[!ht]
   \begin{center}
    \leavevmode
\centerline{\epsfig{file=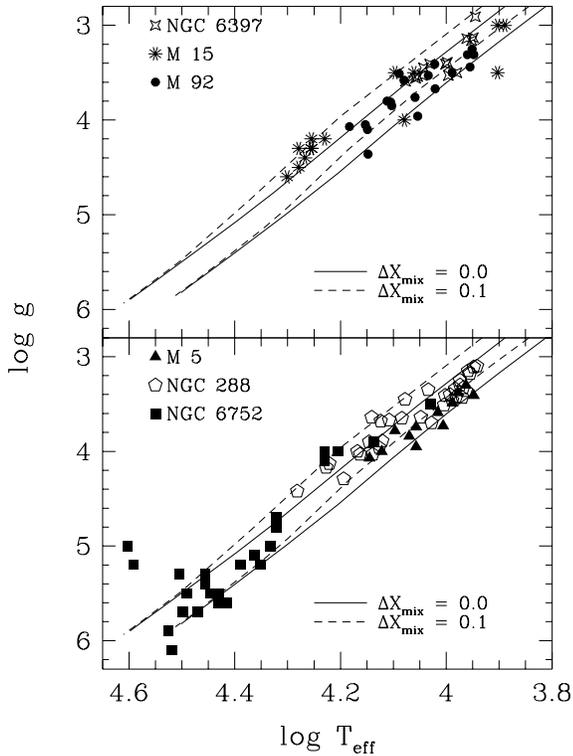,height=8.0cm}}
   \end{center}
\vspace*{1.1cm}
\caption[]{Gravities of BHB stars are compared to the location of the 
HB band from the calculation of Sweigart (1997) without helium mixing
(full drawn lines) and adopting a helium
mixing parameter of $\Delta$X$_{\rm mix}$=0.1 (dashed lines). 
{\bf Top:} Comparison
for the metal-poor clusters NGC\,6397, M\,15 and M\,92.
{\bf Bottom:} Same, but for the intermediate metallicity clusters M\,5, 
NGC\,288 and NGC\,6752.}
\end{figure}
}

\section{ABSOLUTE MAGNITUDES OF BHB STARS REVISITED}

Increasing the distances to the globular clusters also increases the 
absolute visual magnitudes of the Horizontal Branch stars. In Figure 5
we compare the revised absolute magnitudes to the model predictions of 
Sweigart (1997). The theoretical HB band, defined by the zero age position 
(ZAHB) and the 
terminal age HB, is drawn from calculations without helium mixing and with 
a helium mixing parameter of $\Delta$X$_{\rm mix}$=0.1. The metallicity 
adopted in the calculations is Z=0.0005 apropriate for NGC\,6752, but 
slightly too large for the low 
metallicity clusters and too low for M\,5 and NGC\,288. Note, however, that the 
metallicity dependence of the BHB is very small (\cite{dor93}).
The top panel displays the metal poor clusters NGC\,6397,  
M\,15 and M\,92. The observed visual magnitudes are brighter than the
theoretical ZAHB without mixing. The  ZAHB for $\Delta$X$_{\rm mix}$=0.1 gives a resonable 
representation. The bottom panel of Figure 5 displays the more metal rich 
stars. For M\,5 and NGC\,288 
 the observed positions at the cool end (\teff=10000\,K) are well 
reproduced while the slope of the HB at hotter temperatures is shallower 
than predicted. This could possibly be explained if some helium mixing 
occurs at \teff\ $>$ 14000\,K. The sdBs in NGC\,6752 are well reproduced 
by the canonical HB models. Note that the three stars lie above the HB have 
been identified as evolved post-HB stars (\cite{moe97}).

{\small
\begin{figure}[!ht]
   \begin{center}
    \leavevmode
\centerline{\epsfig{file=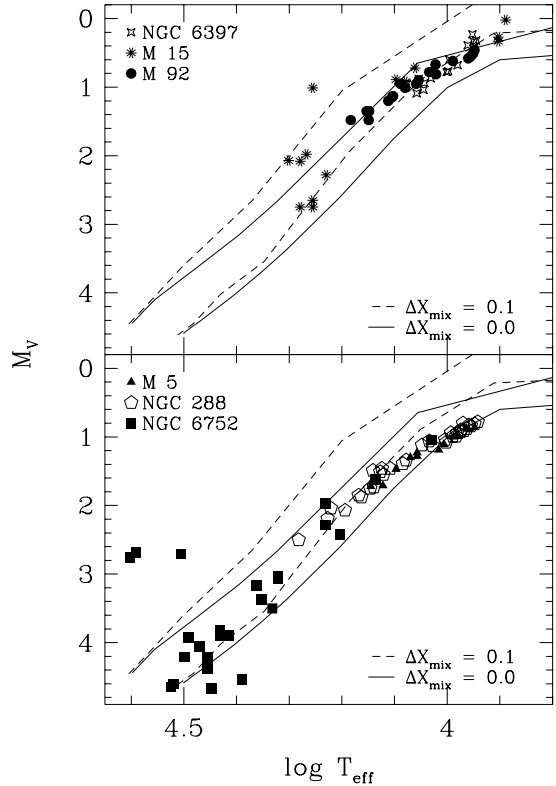,height=8.0cm}}
   \end{center}
\vspace*{1.1cm}
\caption[]{Absolute magnitudes of BHB stars calculated from the 
new distances of Reid (1997a,b) are compared to the location of the 
HB band from the calculation of Sweigart (1997) without helium mixing
(full drawn lines) and adopting a helium
mixing parameter of $\Delta$X$_{\rm mix}$=0.1 (dashed lines). 
{\bf Top:} Comparison
for the metal-poor clusters NGC\,6397, M\,15 and M\,92.
{\bf Bottom:} Same, but for the intermediate metallicity clusters M\,5, 
NGC\,288 and NGC\,6752.}
\end{figure}
}

\section{CONCLUSSIONS}

We have redetermined masses and absolute visual magnitudes 
of BHB stars in 
six globular clusters using new (larger) cluster 
distances derived by Reid (1997a,b) from HIPPARCOS parallaxes of local 
subdwarfs by main sequence fitting.  
The masses for the BHB stars are larger than previous estimates. At the 
new distance scale the masses for NGC\,6752, M\,5, M\,15 and M\,92 are 
consistent with model predictions of canonical evolutionary models.
For NGC\,288 and 
NGC\,6397  the BHB star masses are still  
slightly too low but the mass deficit is reduced. The absolute visual 
magnitudes of BHB stars are consistent with those predicted by canonical theory
for the cool end of the more metal rich clusters M\,5 and NGC\,288 as well 
as for the sdBs in NGC\,6752 whereas 
the hotter BHB stars in these cluster as well as all BHB stars 
in the metal poor clusters are brighter than predicted by canonical 
ZAHB models. 

In summary, the new distances to six globular cluster remove the problem of
too low masses in two cases and weaken it in two other cases. However, 
at the same time they create a new problem for the BHB absolute magnitudes 
being too bright (at least for the metal poor clusters). 
If the new distance scale
proves to be correct, the canonical evolutionary models underestimate the
BHB brightnesses. This conclusion is corroborated by the too low gravities 
found for some parts of the BHB. Hence, we considered non-canonical tracks 
(\cite{swe97}) which included helium mixing and resulted in brighter 
BHB stars with lower gravities. While these models can reproduce the BHB 
magnitudes and gravities, further research is required to clarify whether 
deep He mixing occurs and which parameters determine the amount of mixing.

\end{document}